\RequirePackage{ifpdf}
\ifpdf 
\documentclass[pdftex]{sigma}
\else
\documentclass{sigma}
\fi

\numberwithin{equation}{section}

\begin{document}

\allowdisplaybreaks

\renewcommand{\thefootnote}{$\star$}

\renewcommand{\PaperNumber}{059}

\FirstPageHeading

\ShortArticleName{Exact Solutions for an Ultradiscrete Painlev\'e Equation}

\ArticleName{Exact Solutions with Two Parameters\\ for an Ultradiscrete Painlev\'e Equation of Type $\boldsymbol{A_6^{(1)}}$\,\footnote{This paper is a
contribution to the Special Issue ``Relationship of Orthogonal Polynomials and Special Functions with Quantum Groups and Integrable Systems''. The
full collection is available at
\href{http://www.emis.de/journals/SIGMA/OPSF.html}{http://www.emis.de/journals/SIGMA/OPSF.html}}}

\Author{Mikio MURATA}

\AuthorNameForHeading{M.~Murata}

\Address{Department of Physics and Mathematics, College of Science and Engineering, Aoyama Gakuin University,
5-10-1 Fuchinobe, Chuo-ku, Sagamihara-shi, Kanagawa, 252-5258 Japan}
\Email{\href{mailto:murata@gem.aoyama.ac.jp}{murata@gem.aoyama.ac.jp}}

\ArticleDates{Received February 07, 2011, in f\/inal form June 11, 2011;  Published online June 17, 2011}

\Abstract{An ultradiscrete system corresponding to the $q$-Painlev\'e equation of type $A_6^{(1)}$, which is a $q$-dif\/ference analogue of the second Painlev\'e equation, is proposed.
Exact solutions with two parameters are constructed for the ultradiscrete system.}


\Classification{33E17; 39A12}

\section{Introduction}

Discrete Painlev\'e equations are prototype integrable systems studied from various points of view~\cite{RGH,Sakai2001}.
They are discrete equations which are reduced to the Painlev\'e equations in suitable limiting processes, and moreover, which pass the singularity conf\/inement test \cite{GRP}.
Many results are already given about special solutions of discrete Painlev\'e equations \cite{HKW,KMNOY,KMNOY2,KOSGR,MSY,RGTT}.

Ultradiscretization \cite{TTMS} is a limiting procedure transforming a given dif\/ference equation into a~cellular automaton. In addition the cellular automaton constructed by this procedure preserves the essential properties of the original equation, such as the structure of exact solutions. In this procedure, we f\/irst replace a dependent variable $x_n$ in a given equation by
\begin{gather*}
x_n = \exp\left(\frac{X_n}{\varepsilon}\right),
\end{gather*}
where $\varepsilon$ is a positive parameter. Then, we apply $\varepsilon \log$ to both sides of the equation and take the limit $\varepsilon \to +0$. Using identity
\begin{gather*}
\lim_{\varepsilon \to +0}
 \varepsilon \log\big(e^{X/\varepsilon} + e^{Y/\varepsilon}\big)
= \max\left(X,  Y \right)
\end{gather*}
and exponential laws, we f\/ind that addition, multiplication, and division for the original variables are replaced by maximum, addition, and subtraction for the new ones, respectively. In this way the original dif\/ference equation is approximated to a piecewise linear equation which can be regarded as a time evolution rule for a cellular automaton.

It is an interesting problem to study ultradiscrete analogues of the Painlev\'{e} equations and the structure of their solutions. Some ultradiscrete Painlev\'e equations and their special solutions are studied in, for example, \cite{GORTT,JL2005,JL2006,JNO,Ormerod,RTGO,TTGOR}.
However the structure of the general solutions is completely unclear today.

In this paper we propose a new ultradiscrete Painlev\'e equation of simultaneous type.
With this purpose, we start with a $q$-Painlev\'e equation of type $A_6^{(1)}$ ($q$-$P(A_6)$) \cite{HKW,KMNOY,KMNOY2,NishiokaS,NishiokaS2,Sakai2007,Sakai2001}
  \begin{gather}\label{eqs:qpa6}
  f_nf_{n-1} =1+g_{n-1},\qquad
  g_ng_{n-1} =\frac{a q^{2n}f_n}{f_n+q^n},
  \end{gather}
where $a$ and $q$ are parameters.
Equation \eqref{eqs:qpa6} is the simplest nontrivial $q$-Painlev\'e equation that admits a B\"acklund transformation.
This equation is also referred to as $q$-analogue of the second Painlev\'e equation
  \begin{gather*}
\left(f_{n+1}f_{n}-1\right)\left(f_nf_{n-1}-1\right)
=\frac{a q^{2n}f_n}{f_n+q^n}
  \end{gather*}
and reduced to the second Painlev\'e equation
\begin{gather*}
\frac{d^2 y}{ d s^2} = 2 y^3 + 2 s  y + c 
\end{gather*}
in a continuous limit \cite{RG}.

Furthermore, we propose an exact solution with two parameters for the ultradiscrete system.
Although the Painlev\'e equations and the $q$-analogues of these are not generally solvable in terms of elementary functions \cite{NishiokaK,NishiokaS,NO,Umemura}, it is an amazing fact that the ultradiscrete analogues of these are ``solvable''.

In Section~\ref{sec:qpudp}, we present an ultradiscrete analogue of $q$-$P(A_6)$.
In Section~\ref{sec:sol}, we give an exact solution with two parameters of this ultradiscrete system.
In Section~\ref{sec:bt}, we construct an ultradiscrete B\"acklund transformation.
The exact solutions with two parameters are also obtained from a ``seed'' solution.
In Section~\ref{sec:spe}, we give ultradiscrete hypergeometric solutions which are included in the solutions with two parameters.
Finally concluding remarks are given in Section~\ref{sec:rem}.

\section{Ultradiscrete Painlev\'e equation}\label{sec:qpudp}

We construct an ultradiscrete analogue of $q$-$P(A_6)$ \eqref{eqs:qpa6}.
Let us introduce
\begin{gather*}
  f_n =\exp\left(F_n/\varepsilon\right),\qquad
  g_n =\exp\left(G_n/\varepsilon\right),\qquad
  q =\exp\left(Q/\varepsilon\right),\qquad
  a =\exp\left(A/\varepsilon\right)
  \end{gather*}
and take the limit $\varepsilon\to +0$.
Then $q$-$P(A_6)$ \eqref{eqs:qpa6} is reduced to an ultradiscrete analogue of $q$-$P(A_6)$ (ud-$P(A_6)$),
\begin{subequations}\label{eqs:udpa6}
\begin{gather}
  F_n+F_{n-1} =\max\left(0,G_{n-1}\right),\label{eq:udpa6f}\\
  G_n+G_{n-1} =A+2nQ-\max\left(0,nQ-F_{n}\right).\label{eq:udpa6g}
  \end{gather}
\end{subequations}
Because one cannot make a known second order single equation from this system, this ud-$P(A_6)$ is an essentially new ultradiscrete Painlev\'e system.

In \cite{IKMMS}, we have given another ud-$P(A_6)$ by means of ultradiscretization with parity variables, which is an extended version of ultradiscrete procedure.
This procedure keeps track of the sign of original variables \cite{MIMS}.
We have also presented its special solution
that corresponds to the hypergeometric solution in the discrete system.

\section{Solutions}\label{sec:sol}
In order to construct a solution of ud-$P(A_6)$, we take the following strategy.
First we seek solutions for linear systems which are obtained from the piecewise linear system.
These solutions satisfy ud-$P(A_6)$ in some restricted range of $n$.
Next we connect these solutions together to ensure that they satisfy \eqref{eqs:udpa6} for any $n$.

\begin{theorem}\label{thm:1}
ud-$P(A_6)$ admits the following solution for $Q>0$, $A=2(m+r)Q$, $m\in \mathbb{N}$, $-1/2< r\le 1/2$:
\begin{gather*}
F_n =d_1\left(-1\right)^{n-m},\qquad
G_n =\frac{2n+2m+2r+1}{2}Q+d_2\left(-1\right)^{n-m},
  \end{gather*}
for $n\le -m-1$, where $d_1$ and $d_2$ satisfy
\begin{gather*}
-\left(m+2\right)Q\le d_1\le \left(m+1\right)Q,\qquad
\frac{2r-5}{2}Q\le d_2\le \frac{3-2r}{2}Q;
\\
F_n =\frac{n+m+r}{2}Q+e_1\left(-1\right)^{n-m}-e_2\left(n-m\right)\left(-1\right)^{n-m} ,\\
G_n =\frac{2n+2m+2r+1}{2}Q+e_2\left(-1\right)^{n-m},
  \end{gather*}
for $-m\le n\le m-1$,  where $e_1$ and $e_2$ satisfy
 \begin{gather*}
  -\frac{1+2r}{2}Q\le e_2\le \frac{3+2r}{2}Q, \qquad
   e_1+e_2\le\frac{1+r}{2}Q,\qquad
  e_1+2e_2\ge -\frac{2+r}{2}Q, \\
   e_1+\left(2m-1\right)e_2\le\frac{2m+r-1}{2}Q, \qquad
  e_1+2m  e_2\ge-\frac{2m+r}{2}Q,
 \end{gather*}
and
\begin{gather*}
F_n =\frac{n+2m+2r}{3}Q+h_1\cos\frac{2}{3}\pi \left(n-m\right) +\frac{2h_2-h_1}{\sqrt{3}}\sin\frac{2}{3}\pi \left(n-m\right),\\
G_n =\frac{2n+4m+4r+1}{3}Q+h_2\cos\frac{2}{3}\pi \left(n-m\right) +\frac{h_2-2h_1}{\sqrt{3}}\sin\frac{2}{3}\pi \left(n-m\right),
  \end{gather*}
for $n\ge m$, where $h_1$ and $h_2$ satisfy
 \begin{gather*}
 h_1\le \frac{6-2r}{3}Q,\qquad
 h_2\ge \frac{2r-4}{3}Q, \qquad
 h_2-h_1\le \frac{2-2r}{3}Q.
 \end{gather*}
Here the relations between $d_1$, $d_2$ and $e_1$, $e_2$ are
\begin{gather*}
d_1 =\frac{r}{2}Q+e_1+2m  e_2-2\max\left(0,\frac{2r-1}{2}Q-e_2\right),\qquad
d_2 =e_2,
  \end{gather*}
and those between $e_1$, $e_2$ and $h_1$, $h_2$ are
\begin{gather*}
h_1 =-\frac{r}{6}Q+e_1,\qquad
h_2 =\frac{1-2r}{6}Q+e_2-\max\left(0,-\frac{r}{2}Q-e_1\right).
  \end{gather*}
\end{theorem}

\begin{proof}
We consider the case $A=2(m+r)Q$, $m\in \mathbb{N}$ and $-1/2< r\le 1/2$.
If $G_{n-1}\le 0$ and $n Q-F_{n} \le 0$, then ud-$P(A_6)$ \eqref{eqs:udpa6} can be written as the following system of linear equations:
\begin{gather}\label{eqs:udpa6-1}
  F_n+F_{n-1} =0,\qquad
  G_n+G_{n-1} =\left(2n+2m+2r\right)Q.
  \end{gather}
The general solution to the linear system \eqref{eqs:udpa6-1} is
\begin{gather}\label{eqs:udpa6-1sol}
F_n =d_1\left(-1\right)^{n-m},\qquad
G_n =\frac{2n+2m+2r+1}{2}Q+d_2\left(-1\right)^{n-m},
  \end{gather}
where $d_1$ and $d_2$ are arbitrary constants.
If $d_1=d_2=0$, the particular solution \eqref{eqs:udpa6-1sol} satisf\/ies $G_{n-1}\le 0$ and $n Q-F_{n} \le 0$ for $n\le -m-1$.
The suf\/f\/icient condition that the general solution~\eqref{eqs:udpa6-1sol} satisf\/ies $G_{n-1}\le 0$ and $n Q-F_{n} \le 0$ for $n\le -m-1$ is
\begin{gather}\label{eqs:udpa6-1con}
-\left(m+2\right)Q\le d_1\le \left(m+1\right)Q,\qquad
\frac{2r-5}{2}Q\le d_2\le \frac{3-2r}{2}Q.
 \end{gather}
Therefore \eqref{eqs:udpa6-1sol} that satisf\/ies \eqref{eqs:udpa6-1con} is a solution to ud-$P(A_6)$ for $n\le -m-1$.
If $G_{n-1}\ge 0$ and $n Q-F_{n} \le 0$, then ud-$P(A_6)$ \eqref{eqs:udpa6} can be written as the following system of linear equations:
\begin{gather}\label{eqs:udpa6-2}
  F_n+F_{n-1} =G_{n-1},\qquad
  G_n+G_{n-1} =\left(2n+2m+2r\right)Q.
  \end{gather}
The general solution to the linear system \eqref{eqs:udpa6-2} is
\begin{gather}
F_n =\frac{n+m+r}{2}Q+e_1\left(-1\right)^{n-m}-e_2\left(n-m\right)\left(-1\right)^{n-m} ,\nonumber\\
G_n =\frac{2n+2m+2r+1}{2}Q+e_2\left(-1\right)^{n-m},\label{eqs:udpa6-2sol}
  \end{gather}
where $e_1$ and $e_2$ are arbitrary constants.
If $e_1=e_2=0$, \eqref{eqs:udpa6-2sol} satisf\/ies $G_{n}\ge 0$ and $n Q-F_{n} \le 0$ for $-m\le n\le m-1$.
The condition that the general solution~\eqref{eqs:udpa6-2sol} satisf\/ies $G_{n}\ge 0$ and $n Q-F_{n} \le 0$ for $-m\le n\le m-1$ is
 \begin{gather}
 -\frac{1+2r}{2}Q\le e_2\le \frac{3+2r}{2}Q, \qquad
 e_1+e_2\le\frac{1+r}{2}Q, \qquad
e_1+2e_2\ge -\frac{2+r}{2}Q,\nonumber\\
  e_1+\left(2m-1\right)e_2\le\frac{2m+r-1}{2}Q,\qquad
e_1+2m  e_2\ge-\frac{2m+r}{2}Q.\label{eqs:udpa6-2con}
 \end{gather}
Therefore \eqref{eqs:udpa6-2sol} that satisf\/ies \eqref{eqs:udpa6-2con} is a solution to ud-$P(A_6)$ for $-m\le n\le m-1$.
If $G_{n-1}\ge 0$ and $n Q-F_{n} \ge 0$, then ud-$P(A_6)$ \eqref{eqs:udpa6} can be written as the following system of linear equations:
\begin{gather}
  F_n+F_{n-1} =G_{n-1},\qquad
  G_n+G_{n-1} =\left(n+2m+2r\right)Q+F_{n}.\label{eqs:udpa6-3}
  \end{gather}
The general solution to the linear system \eqref{eqs:udpa6-3} is
\begin{gather}
F_n =\frac{n+2m+2r}{3}Q+h_1\cos\frac{2}{3}\pi \left(n-m\right) +\frac{2h_2-h_1}{\sqrt{3}}\sin\frac{2}{3}\pi \left(n-m\right),\nonumber\\
G_n =\frac{2n+4m+4r+1}{3}Q+h_2\cos\frac{2}{3}\pi \left(n-m\right) +\frac{h_2-2h_1}{\sqrt{3}}\sin\frac{2}{3}\pi \left(n-m\right),\label{eqs:udpa6-3sol}
  \end{gather}
where $h_1$ and $h_2$ are arbitrary constants.
If $h_1=h_2=0$, \eqref{eqs:udpa6-3sol} satisf\/ies $G_{n-1}\ge 0$ and $n Q-F_{n} \ge 0$ for $n\ge m+1$.
The condition that the general solution~\eqref{eqs:udpa6-3sol} satisf\/ies $G_{n-1}\ge 0$ and $n Q-F_{n} \ge 0$ for $n\ge m+1$ is
\begin{gather}
 h_1\le \frac{6-2r}{3}Q,\qquad
 h_2\ge \frac{2r-4}{3}Q, \qquad
 h_2-h_1\le \frac{2-2r}{3}Q.\label{eqs:udpa6-3con}
 \end{gather}
Therefore \eqref{eqs:udpa6-3sol} that satisf\/ies \eqref{eqs:udpa6-3con} is a solution to ud-$P(A_6)$ for $n\ge m+1$.
The relations between $d_1$, $d_2$ and $e_1$, $e_2$ can be obtained from
\eqref{eq:udpa6f} for $n=-m$:
 \begin{gather*}
 F_{-m}+F_{-m-1} =\max\left(0, G_{-m-1}\right),
 \end{gather*}
\eqref{eqs:udpa6-1sol} for $n=-m-1$:
\begin{gather*}
 F_{-m-1} =-d_1,\qquad
G_{-m-1} =\frac{2r-1}{2}Q-d_2,
 \end{gather*}
and \eqref{eqs:udpa6-2sol} for $n=-m, -m-1$ respectively:
\begin{gather*}
 F_{-m} =\frac{r}{2}Q+2m  e_2+e_1, \qquad
G_{-m-1} =\frac{2r-1}{2}Q-e_2.
 \end{gather*}
We have
\begin{gather*}
d_1 =\frac{r}{2}Q+e_1+2m  e_2-2\max\left(0, \frac{2r-1}{2}Q-e_2\right),\qquad
d_2 =e_2.
  \end{gather*}
Moreover the relations between $e_1$, $e_2$ and $h_1$, $h_2$ can be obtained from
\eqref{eq:udpa6g} for $n=m$:
\begin{gather*}
 G_m+G_{m-1} =\left(4m+2r\right)Q-\max\left(0, m Q-F_{m}\right),
  \end{gather*}
\eqref{eqs:udpa6-2sol} for $n=m, m-1$ respectively:
  \begin{gather*}
F_m =\frac{2m+r}{2}Q+e_1,\qquad
G_{m-1} =\frac{4m+2r-1}{2}Q-e_2,
  \end{gather*}
and \eqref{eqs:udpa6-3sol} for $n=m$:
\begin{gather*}
F_m =\frac{3m+2r}{3}Q+h_1,\qquad
G_m =\frac{6m+4r+1}{3}Q+h_2.
  \end{gather*}
And we have
\begin{gather*}
h_1 =-\frac{r}{6}Q+e_1,\qquad
h_2 =\frac{1-2r}{6}Q+e_2-\max\left(0, -\frac{r}{2}Q-e_1\right).
  \end{gather*}
When $|e_1|$ and $|e_2|$ are suf\/f\/iciently small,
we shall write ``$e_1\sim 0$, $e_2\sim 0$'' as an abbreviation,
If $e_1\sim 0$ and $e_2\sim 0$, then we f\/ind that
\begin{gather*}
d_1 \sim \frac{r}{2}Q,\qquad
d_2 \sim 0
  \end{gather*}
satisfy \eqref{eqs:udpa6-1con},
and
\begin{gather*}
h_1 \sim -\frac{r}{6}Q,\qquad
h_2 \sim \frac{1-2r}{6}Q-\max\left(0, -\frac{r}{2}Q\right)
  \end{gather*}
satisfy \eqref{eqs:udpa6-3con}.
Therefore
we have Theorem~\ref{thm:1} by connecting these solutions together.
\end{proof}

\begin{theorem}\label{thm:2}
ud-$P(A_6)$ admits the following solution for $Q>0$, $A=2(m+r)Q$, $-m\in \mathbb{N}$, $0< r\le 1/2$:
\begin{gather*}
F_n =d_1\left(-1\right)^{n},\qquad
G_n =\frac{2n+2m+2r+1}{2}Q+d_2\left(-1\right)^{n}
  \end{gather*}
for $n\le -1$, where $d_1$ and $d_2$ satisfy
\begin{gather*}
-2Q\le d_1\le Q,\qquad
\frac{2m+2r-1}{2}Q\le d_2\le \frac{-2m-2r+3}{2}Q;
\\
F_n =e_1\left(-1\right)^n,\qquad
G_n =\frac{2n+4m+4r+1}{4}Q+e_1n\left(-1\right)^n+e_2\left(-1\right)^n
  \end{gather*}
for $0\le n\le -2m-1$, where $e_1$ and $e_2$ satisfy
\begin{gather*}
-Q\le e_1\le 2Q,\qquad
e_2\le -\frac{4m+4r+1}{4}Q,\qquad
e_1+e_2\ge \frac{4m+4r+3}{4}Q,\\
-\left(2m+2\right)e_1+e_2\le \frac{3-4r}{4}Q, \qquad
-\left(2m+3\right)e_1+e_2\ge \frac{4r-5}{4}Q,
 \end{gather*}
and
\begin{gather*}
F_n =\frac{n+2m+2r}{3}Q+h_1\cos\frac{2}{3}\pi \left(n+2m\right) +\frac{2h_2-h_1}{\sqrt{3}}\sin\frac{2}{3}\pi \left(n+2m\right),\\
G_n =\frac{2n+4m+4r+1}{3}Q+h_2\cos\frac{2}{3}\pi \left(n+2m\right) +\frac{h_2-2h_1}{\sqrt{3}}\sin\frac{2}{3}\pi \left(n+2m\right)
  \end{gather*}
for $n\ge -2m$, where $h_1$ and $h_2$ satisfy
\begin{gather*}
 h_1\le \frac{4r+3}{3}Q,\qquad
 h_2\ge -\frac{4r+1}{3}Q, \qquad
 h_2-h_1\le \frac{4r+5}{3}Q.
 \end{gather*}
Here the relations between $d_1$, $d_2$ and $e_1$, $e_2$ are
\begin{gather*}
d_1 =e_1,\qquad
d_2 =-\frac{1}{4}Q+e_2+\max\left(0, -e_1\right),
  \end{gather*}
and those between $e_1$, $e_2$ and $h_1$, $h_2$ are
\begin{gather*}
h_1 =-\frac{2r}{3}Q+e_1+\max\left\{0, \frac{4r-1}{4}Q+\left(2m+1\right)e_1-e_2\right\},\\
h_2 =-\frac{4r+1}{12}Q-2m  e_1+e_2+\max\left\{0, \frac{4r-1}{4}Q+\left(2m+1\right)e_1-e_2\right\}.
  \end{gather*}
\end{theorem}

\begin{theorem}\label{thm:3}
ud-$P(A_6)$  admits the following solution for $Q>0$, $A=2(m+r)Q$, $-m\in \mathbb{N}$, $-1/2< r\le 0$:
\begin{gather*}
F_n =d_1\left(-1\right)^{n},\qquad
G_n =\frac{2n+2m+2r+1}{2}Q+d_2\left(-1\right)^{n}
  \end{gather*}
for $n\le -1$, where $d_1$ and $d_2$ satisfy
\begin{gather*}
-2Q\le d_1\le Q,\qquad
\frac{2m+2r-1}{2}Q\le d_2\le \frac{-2m-2r+3}{2}Q;
 \\
F_n =e_1\left(-1\right)^n,\qquad
G_n =\frac{2n+4m+4r+1}{4}Q+e_1n\left(-1\right)^n+e_2\left(-1\right)^n
\end{gather*}
for $0\le n\le -2m$, where $e_1$ and $e_2$ satisfy
\begin{gather*}
-Q\le e_1\le 2Q,\qquad
e_2\le -\frac{4m+4r+1}{4}Q,\qquad
e_1+e_2\ge \frac{4m+4r+3}{4}Q,\\
-\left(2m+1\right)e_1+e_2\ge \frac{4r-1}{4}Q,\qquad
-\left(2m+2\right)e_1+e_2\le \frac{3-4r}{4}Q,
 \end{gather*}
and
\begin{gather*}
F_n =\frac{n+2m+2r}{3}Q+h_1\cos\frac{2}{3}\pi \left(n+2m\right) +\frac{2h_2-h_1}{\sqrt{3}}\sin\frac{2}{3}\pi \left(n+2m\right),\\
G_n =\frac{2n+4m+4r+1}{3}Q+h_2\cos\frac{2}{3}\pi \left(n+2m\right) +\frac{h_2-2h_1}{\sqrt{3}}\sin\frac{2}{3}\pi \left(n+2m\right)
  \end{gather*}
for $n\ge -2m+1$, where $h_1$ and $h_2$ satisfy
\begin{gather*}
 h_1\le \frac{4r+3}{3}Q,\qquad
 h_2\ge -\frac{4r+7}{3}Q, \qquad
 h_2-h_1\le \frac{4r+5}{3}Q.
 \end{gather*}
Here the relations between $d_1$, $d_2$ and $e_1$, $e_2$ are
\begin{gather*}
d_1 =e_1,\qquad
d_2 =-\frac{1}{4}Q+e_2+\max\left(0, -e_1\right),
  \end{gather*}
and those between $e_1$, $e_2$ and $h_1$, $h_2$ are
\begin{gather*}
h_1 =\frac{4r+3}{12}Q-\left(2m-1\right)e_1+e_2-\max\left(0, \frac{4r+1}{4}Q-2m  e_1+e_2\right),\\
h_2 =-\frac{4r+1}{12}Q-2m  e_1+e_2.
  \end{gather*}
\end{theorem}

\begin{proof}
We consider the case $A=2(m+r)Q$, $-m\in \mathbb{N}$ and $-1/2< r\le 1/2$.
If $G_{n-1}\le 0$ and $n Q-F_{n} \le 0$, then ud-$P(A_6)$ \eqref{eqs:udpa6} can be written as the following system of linear equations:
\begin{gather}
  F_n+F_{n-1} =0,\qquad
  G_n+G_{n-1} =\left(2n+2m+2r\right)Q.\label{eqs:udpa6-n1}
  \end{gather}
The general solution to the linear system \eqref{eqs:udpa6-n1} is
\begin{gather}
F_n =d_1\left(-1\right)^{n},\qquad
G_n =\frac{2n+2m+2r+1}{2}Q+d_2\left(-1\right)^{n},\label{eqs:udpa6-n1sol}
  \end{gather}
where $d_1$ and $d_2$ are arbitrary constants.
If $d_1=d_2=0$, the particular solution~\eqref{eqs:udpa6-n1sol} satisf\/ies $G_{n}\le 0$ and $n Q-F_{n} \le 0$ for $n\le -1$.
The condition that the general solution~\eqref{eqs:udpa6-n1sol} satisf\/ies $G_{n-1}\le 0$ and $n Q-F_{n} \le 0$ for $n\le -1$ is
\begin{gather}
-2Q\le d_1\le Q,\qquad
\frac{2m+2r-1}{2}Q\le d_2\le \frac{-2m-2r+3}{2}Q.\label{eqs:udpa6-n1con}
 \end{gather}
Therefore \eqref{eqs:udpa6-n1sol} that satisf\/ies~\eqref{eqs:udpa6-n1con} is a solution to ud-$P(A_6)$ for $n\le -1$.
If $G_{n-1}\le 0$ and $n Q-F_{n} \ge 0$, then~\eqref{eqs:udpa6} can be written as the following system of linear equations:
\begin{gather}
  F_n+F_{n-1} =0,\qquad
  G_n+G_{n-1} =\left(n+2m+2r\right)Q+F_{n}.\label{eqs:udpa6-n2}
  \end{gather}
The general solution to the linear system \eqref{eqs:udpa6-n2} is
\begin{gather}
F_n =e_1\left(-1\right)^n,\qquad
G_n =\frac{2n+4m+4r+1}{4}Q+e_1n\left(-1\right)^n+e_2\left(-1\right)^n,\label{eqs:udpa6-n2sol}
  \end{gather}
where $e_1$ and $e_2$ are arbitrary constants.
If $e_1=e_2=0$ and $0<r\le 1/2$, \eqref{eqs:udpa6-n2sol} satisf\/ies $G_{n-1}\le 0$ and $n Q-F_{n} \ge 0$ for $1\le n\le -2m-1$.
The condition that the general solution~\eqref{eqs:udpa6-n2sol} satisf\/ies $G_{n-1}\le 0$ and $n Q-F_{n} \ge 0$ for $1\le n\le -2m-1$ is
\begin{gather}
 -Q\le e_1\le 2Q,\qquad
e_2\le -\frac{4m+4r+1}{4}Q,\qquad
e_1+e_2\ge \frac{4m+4r+3}{4}Q,\nonumber\\
-\left(2m+2\right)e_1+e_2\le \frac{3-4r}{4}Q,\qquad
-\left(2m+3\right)e_1+e_2\ge \frac{4r-5}{4}Q.\label{eqs:udpa6-n2con}
 \end{gather}
Therefore \eqref{eqs:udpa6-n2sol} that satisf\/ies \eqref{eqs:udpa6-n2con} is a solution to ud-$P(A_6)$ for $1\le n\le-2 m-1$.
If $e_1=e_2=0$ and $-1/2<r\le 0$, then~\eqref{eqs:udpa6-n2sol} satisf\/ies $G_{n-1}\le 0$ and $n Q-F_{n} \ge 0$ for $1\le n\le -2m$.
The condition that the general solution~\eqref{eqs:udpa6-n2sol} satisf\/ies $G_{n-1}\le 0$ and $n Q-F_{n} \ge 0$ for $1\le n\le -2m$ is
\begin{gather}
 -Q\le e_1\le 2Q,\qquad
 e_2\le -\frac{4m+4r+1}{4}Q,\qquad
 e_1+e_2\ge \frac{4m+4r+3}{4}Q,\nonumber\\
 -\left(2m+1\right)e_1+e_2\ge \frac{4r-1}{4}Q,\qquad
-\left(2m+2\right)e_1+e_2\le \frac{3-4r}{4}Q.\label{eqs:udpa6-n2con2}
\end{gather}
Therefore \eqref{eqs:udpa6-n2sol} that satisf\/ies \eqref{eqs:udpa6-n2con2} is a solution to ud-$P(A_6)$ for $1\le n\le-2 m$.
If $G_{n-1}\ge 0$ and $n Q-F_{n} \ge 0$, then ud-$P(A_6)$ \eqref{eqs:udpa6} can be written as the following system of linear equations:
\begin{gather}
  F_n+F_{n-1} =G_{n-1},\qquad
  G_n+G_{n-1} =\left(n+2m+2r\right)Q+F_{n}.\label{eqs:udpa6-n3}
  \end{gather}
The general solution to the linear system \eqref{eqs:udpa6-n3} is
\begin{gather}
F_n =\frac{n+2m+2r}{3}Q+h_1\cos\frac{2}{3}\pi \left(n+2m\right) +\frac{2h_2-h_1}{\sqrt{3}}\sin\frac{2}{3}\pi \left(n+2m\right),\nonumber\\
G_n =\frac{2n+4m+4r+1}{3}Q+h_2\cos\frac{2}{3}\pi \left(n+2m\right) +\frac{h_2-2h_1}{\sqrt{3}}\sin\frac{2}{3}\pi \left(n+2m\right),\label{eqs:udpa6-n3sol}
  \end{gather}
where $h_1$ and $h_2$ are arbitrary constants.
If $h_1=h_2=0$ and $0<r\le 1/2$, \eqref{eqs:udpa6-n3sol} satisf\/ies the conditions $G_{n}\ge 0$ and $n Q-F_{n} \ge 0$ for $n\ge -2m$.
The condition that the general solution~\eqref{eqs:udpa6-n3sol} satisf\/ies $G_{n}\ge 0$ and $n Q-F_{n} \ge 0$ for $n\ge -2m$ is
\begin{gather}
 h_1\le \frac{4r+3}{3}Q,\qquad
 h_2\ge -\frac{4r+1}{3}Q, \qquad
 h_2-h_1\le \frac{4r+5}{3}Q.\label{eqs:udpa6-n3con}
 \end{gather}
Therefore \eqref{eqs:udpa6-n3sol} that satisf\/ies \eqref{eqs:udpa6-n3con} is a solution to ud-$P(A_6)$ for $n\ge -2m$.
If $h_1=h_2=0$ and $-1/2<r\le 0$, \eqref{eqs:udpa6-n3sol} satisf\/ies $G_{n}\ge 0$ and $n Q-F_{n} \ge 0$ for $n\ge -2m+1$.
The condition that the general solution~\eqref{eqs:udpa6-n3sol} satisf\/ies $G_{n}\ge 0$ and $n Q-F_{n} \ge 0$ for $n\ge -2m+1$ is
\begin{gather}
 h_1\le \frac{4r+3}{3}Q,\qquad
 h_2\ge -\frac{4r+7}{3}Q, \qquad
 h_2-h_1\le \frac{4r+5}{3}Q.\label{eqs:udpa6-n3con2}
 \end{gather}
Therefore \eqref{eqs:udpa6-n3sol} that satisf\/ies \eqref{eqs:udpa6-n3con2} is a solution to ud-$P(A_6)$ for $n\ge -2m+1$.
The relations between $d_1$, $d_2$ and $e_1$, $e_2$  can be obtained from
\eqref{eq:udpa6g} for $n=0$:
\begin{gather*}
  G_0+G_{-1} =\left(2m+2r\right)Q-\max\left(0, -F_{0}\right),
\end{gather*}
\eqref{eqs:udpa6-n1sol} for $n=0, -1$ respectively:
\begin{gather*}
F_0 =d_1,\qquad
G_{-1} =\frac{2m+2r-1}{2}Q-d_2,
  \end{gather*}
and \eqref{eqs:udpa6-n2sol} for $n=0$:
\begin{gather*}
F_0 =e_1,\qquad
G_0 =\frac{4m+4r+1}{4}Q+e_2.
  \end{gather*}
We have
\begin{gather*}
d_1 =e_1,\qquad
d_2 =-\frac{1}{4}Q+e_2+\max\left(0, -e_1\right).
  \end{gather*}
Moreover in the case $0<r\le 1/2$,
the relations between $e_1$, $e_2$ and $h_1$, $h_2$ can be obtained from~\eqref{eq:udpa6f} for $n=-2m$:
 \begin{gather*}
 F_{-2m}+F_{-2m-1} =\max\left(0, G_{-2m-1}\right),
  \end{gather*}
\eqref{eqs:udpa6-n2sol} for $n=-2m-1$:
\begin{gather*}
F_{-2m-1} =-e_1,\qquad
G_{-2m-1} =\frac{4r-1}{4}Q+\left(2m+1\right)e_1-e_2,
  \end{gather*}
and \eqref{eqs:udpa6-n3sol} for $n=-2m,-2m-1$ respectively:
\begin{gather*}
F_{-2m} =\frac{2r}{3}Q+h_1, \qquad
G_{-2m-1} =\frac{4r-1}{3}Q+h_1-h_2.
  \end{gather*}
We have
\begin{gather*}
h_1 =-\frac{2r}{3}Q+e_1+\max\left\{0, \frac{4r-1}{4}Q+\left(2m+1\right)e_1-e_2\right\},\\
h_2 =-\frac{4r+1}{12}Q-2m  e_1+e_2+\max\left\{0, \frac{4r-1}{4}Q+\left(2m+1\right)e_1-e_2\right\}.
  \end{gather*}
In the case $-1/2<r\le 0$,
the relations between $e_1$, $e_2$ and $h_1$, $h_2$ can be obtained from~\eqref{eq:udpa6f} for $n=-2m+1$:
 \begin{gather*}
 F_{-2m+1}+F_{-2m} =\max\left(0, G_{-2m}\right),
  \end{gather*}
\eqref{eqs:udpa6-n2sol} for $n=-2m$:
\begin{gather*}
F_{-2m} =e_1,\qquad
G_{-2m} =\frac{4r+1}{4}Q-2m  e_1+e_2,
  \end{gather*}
and \eqref{eqs:udpa6-n3sol} for $n=-2m+1, -2m$ respectively:
\begin{gather*}
F_{-2m+1} =\frac{2r+1}{3}Q-h_1+h_2,\qquad
G_{-2m} =\frac{4r+1}{3}Q+h_2.
  \end{gather*}
We have
\begin{gather*}
h_1 =\frac{4r+3}{12}Q-\left(2m-1\right)e_1+e_2-\max\left(0, \frac{4r+1}{4}Q-2m  e_1+e_2\right),\\
h_2 =-\frac{4r+1}{12}Q-2m  e_1+e_2.
  \end{gather*}
If $e_1\sim 0$, $e_2\sim 0$,
then we f\/ind that
\begin{gather*}
d_1 \sim 0,\qquad
d_2 \sim -\frac{1}{4}Q
  \end{gather*}
satisfy \eqref{eqs:udpa6-n1con},
\begin{gather*}
h_1 \sim -\frac{2r}{3}Q+\max\left(0, \frac{4r-1}{4}Q\right),\qquad
h_2 \sim -\frac{4r+1}{12}Q+\max\left(0, \frac{4r-1}{4}Q\right)
  \end{gather*}
satisfy \eqref{eqs:udpa6-n3con},
and
\begin{gather*}
h_1 \sim \frac{4r+3}{12}Q-\max\left(0, \frac{4r+1}{4}Q\right),\qquad
h_2 \sim -\frac{4r+1}{12}Q
  \end{gather*}
satisfy \eqref{eqs:udpa6-n3con2}.
We have Theorem~\ref{thm:2} and Theorem~\ref{thm:3} by connecting these solutions together.
\end{proof}

\begin{theorem}\label{thm:0}
ud-$P(A_6)$ admits the following solution for $Q>0$, $A=2r Q$, $-1/2< r\le 1/2$:
\begin{gather*}
F_n =d_1\left(-1\right)^{n},\qquad
G_n =\frac{2n+2r+1}{2}Q+d_2\left(-1\right)^{n},
  \end{gather*}
for $n\le -1$, where $d_1$ and $d_2$ satisfy
\begin{gather*}
-2Q\le d_1\le Q,\qquad
\frac{2r-5}{2}Q\le d_2\le \frac{3-2r}{2}Q,
 \end{gather*}
and
\begin{gather*}
F_n =\frac{n+2r}{3}Q+h_1\cos\frac{2}{3}\pi  n +\frac{2h_2-h_1}{\sqrt{3}}\sin\frac{2}{3}\pi  n,\\
G_n =\frac{2n+4r+1}{3}Q+h_2\cos\frac{2}{3}\pi  n +\frac{h_2-2h_1}{\sqrt{3}}\sin\frac{2}{3}\pi  n,
  \end{gather*}
for $n\ge 1$, where $h_1$ and $h_2$ satisfy
 \begin{gather*}
 h_1\le \frac{4r+3}{3}Q,\qquad
 h_2\ge \frac{2r-4}{3}Q, \qquad
 h_2-h_1\le \frac{2-2r}{3}Q.
 \end{gather*}
Here the relations between $d_1$, $d_2$ and $F_0$, $G_0$ are
\begin{gather*}
d_1 =F_0-\max\left\{0, 2r Q-G_0-\max\left(0, -F_0\right)\right\},\qquad
d_2 =-\frac{2r+1}{2}Q+G_0+\max\left(0, -F_0\right),
  \end{gather*}
and those between $h_1$, $h_2$ and $F_0$, $G_0$ are
\begin{gather*}
h_1 =-\frac{2r}{3}Q+F_0-\max\left(0, -G_0\right),\qquad
h_2 =G_0-\frac{4r+1}{3}Q.
\end{gather*}
\end{theorem}

\begin{proof}
We consider the case $A=2r Q$ and $-1/2< r\le 1/2$.
If $G_{n-1}\le 0$ and $n Q-F_{n} \le 0$, then ud-$P(A_6)$ \eqref{eqs:udpa6} can be written as the following system of linear equations:
\begin{gather}
  F_n+F_{n-1} =0,\qquad
  G_n+G_{n-1} =\left(2n+2r\right)Q.\label{eqs:udpa6-o1}
  \end{gather}
The general solution to the linear system \eqref{eqs:udpa6-o1} is
\begin{gather}
F_n =d_1\left(-1\right)^{n},\qquad
G_n =\frac{2n+2r+1}{2}Q+d_2\left(-1\right)^{n},\label{eqs:udpa6-o1sol}
  \end{gather}
where $d_1$ and $d_2$ are arbitrary constants.
If $d_1=d_2=0$, the particular solution \eqref{eqs:udpa6-o1sol} satisf\/ies $G_{n-1}\le 0$ and $n Q-F_{n} \le 0$ for $n\le -1$.
The suf\/f\/icient condition that the general solution~\eqref{eqs:udpa6-o1sol} satisf\/ies $G_{n-1}\le 0$ and $n Q-F_{n} \le 0$ for $n\le -1$ is
\begin{gather}
-2Q\le d_1\le Q,\qquad
\frac{2r-5}{2}Q\le d_2\le \frac{3-2r}{2}Q.\label{eqs:udpa6-o1con}
 \end{gather}
Therefore \eqref{eqs:udpa6-o1sol} that satisf\/ies \eqref{eqs:udpa6-o1con} is a solution to ud-$P(A_6)$ for $n\le -1$.
If $G_{n-1}\ge 0$ and $n Q-F_{n} \ge 0$, then ud-$P(A_6)$ \eqref{eqs:udpa6} can be written as the following system of linear equations:
\begin{gather}
  F_n+F_{n-1} =G_{n-1},\qquad
  G_n+G_{n-1} =\left(n+2r\right)Q+F_{n}.\label{eqs:udpa6-o3}
  \end{gather}
The general solution to the linear system \eqref{eqs:udpa6-o3} is
\begin{gather}
F_n =\frac{n+2r}{3}Q+h_1\cos\frac{2}{3}\pi  n +\frac{2h_2-h_1}{\sqrt{3}}\sin\frac{2}{3}\pi  n,\nonumber\\
G_n =\frac{2n+4r+1}{3}Q+h_2\cos\frac{2}{3}\pi  n +\frac{h_2-2h_1}{\sqrt{3}}\sin\frac{2}{3}\pi  n,\label{eqs:udpa6-o3sol}
  \end{gather}
where $h_1$ and $h_2$ are arbitrary constants.
If $h_1=h_2=0$, \eqref{eqs:udpa6-o3sol} satisf\/ies $G_{n}\ge 0$ and $n Q-F_{n} \ge 0$ for $n\ge 1$.
The condition that the general solution~\eqref{eqs:udpa6-o3sol} satisf\/ies $G_{n}\ge 0$ and $n Q-F_{n} \ge 0$ for $n\ge 1$ is
\begin{gather}
h_1\le \frac{4r+3}{3}Q,\qquad
h_2\ge \frac{2r-4}{3}Q,\qquad
h_2-h_1\le \frac{2-2r}{3}Q.\label{eqs:udpa6-o3con}
\end{gather}
Therefore \eqref{eqs:udpa6-o3sol} that satisf\/ies \eqref{eqs:udpa6-o3con} is a solution to ud-$P(A_6)$ for $n\ge 2$.
The relations between $d_1$, $d_2$ and $F_0$, $G_0$ can be obtained from
\eqref{eqs:udpa6} for $n=0$:
\begin{gather*}
 F_{0}+F_{-1} =\max\left(0, G_{-1}\right),\qquad
 G_0+G_{-1} =2r Q-\max\left(0, -F_0\right),
 \end{gather*}
and \eqref{eqs:udpa6-o1sol} for $n=-1$:
\begin{gather*}
 F_{-1} =-d_1,\qquad
G_{-1} =\frac{2r-1}{2}Q-d_2.
 \end{gather*}
We have
\begin{gather*}
d_1 =F_0-\max\left\{0, 2r Q-G_0-\max\left(0, -F_0\right)\right\},\qquad
d_2 =-\frac{2r+1}{2}Q+G_0+\max\left(0, -F_0\right).
  \end{gather*}
Moreover the relations between $h_1$, $h_2$ and $F_0$, $G_0$ can be obtained from
\eqref{eq:udpa6f} for $n=1$:
\begin{gather*}
 F_{1}+F_{0} =\max\left(0, G_{0}\right),
  \end{gather*}
and \eqref{eqs:udpa6-o3sol} for $n=1, 0$ respectively:
\begin{gather*}
F_1 =\frac{2r+1}{3}Q-h_1+h_2,\qquad
G_0 =\frac{4r+1}{3}Q+h_2.
  \end{gather*}
And we have
\begin{gather*}
h_1 =-\frac{2r}{3}Q+F_0-\max\left(0, -G_0\right),\qquad
h_2 =G_0-\frac{4r+1}{3}Q.
  \end{gather*}
If $F_0\sim 0$ and $G_0\sim 0$, then we f\/ind that
\begin{gather*}
d_1 \sim -\max\left(0, 2r Q\right),\qquad
d_2 \sim -\frac{2r+1}{2}Q
  \end{gather*}
satisfy \eqref{eqs:udpa6-o1con},
and
\begin{gather*}
h_1 \sim -\frac{2r}{3}Q,\qquad
h_2 \sim -\frac{4r+1}{3}Q
  \end{gather*}
satisfy \eqref{eqs:udpa6-o3con}.
Therefore
we have Theorem~\ref{thm:0} by connecting these solutions together.
\end{proof}

The exact solutions with two parameters for any parameter $A$ have been given in this section.

\section{B\"acklund transformation}\label{sec:bt}

$q$-$P(A_6)$ have the B\"acklund transformation \cite{HKW,Sakai2001}.
That is, if $f_{n}$ and
$g_n$ satisfy $q$-$P(A_6)$ \eqref{eqs:qpa6}, then
\begin{gather}
\mathsf{f}_n =\frac{q^n}{g_n}\frac{a q^{n+1}f_{n+1}+g_n}{q^nf_{n+1}+g_n},\qquad
\mathsf{g}_n =\frac{q^{n+1}}{f_{n+1}}\frac{a q^{n+1}f_{n+1}+g_n}{q^nf_{n+1}+g_n}\label{eqs:bt1}
\end{gather}
satisfy $q$-$P(A_6)$:
\begin{gather*}
  \mathsf{f}_n \mathsf{f}_{n-1} =1+\mathsf{g}_{n-1},\qquad
  \mathsf{g}_n \mathsf{g}_{n-1}
   =\frac{a q^2q^{2n}\mathsf{f}_n}{\mathsf{f}_n+q^n},
  \end{gather*}
and
\begin{gather}
\mathtt{f}_{n+1} =\frac{q^{n+1}}{g_n}\frac{a q^nf_n+g_n}{q^{n+1}f_n+g_n},\qquad
\mathtt{g}_{n} =\frac{q^n}{f_n}\frac{a q^nf_n+g_n}{q^{n+1}f_n+g_n}\label{eqs:bt2}
\end{gather}
also satisfy $q$-$P(A_6)$:
\begin{gather*}
  \mathtt{f}_n \mathtt{f}_{n-1} =1+\mathtt{g}_{n-1},\qquad
  \mathtt{g}_n \mathtt{g}_{n-1}
   =\frac{a q^{-2}q^{2n}\mathtt{f}_n}{\mathtt{f}_n+q^n}.
  \end{gather*}
So we apply the procedure of the ultradiscretization to \eqref{eqs:bt1} and \eqref{eqs:bt2}. Then we have the following theorems.

\begin{theorem}\label{thm:bt1}
If $F_{n}$ and
$G_n$ satisfy ud-$P(A_6)$ \eqref{eqs:udpa6}, then
\begin{gather*}
\mathsf{F}_n =\max\left\{F_{n+1}+\left(n+1\right)Q+A-G_n, 0\right\}
-\max\left(F_{n+1}, G_n-n Q\right),\\
\mathsf{G}_n =Q+\max\left\{\left(n+1\right)Q+A, G_n-F_{n+1}\right\}
-\max\left(F_{n+1}, G_n-n Q\right)
\end{gather*}
satisfy ud-$P(A_6)$:
\begin{gather*}
  \mathsf{F}_n+\mathsf{F}_{n-1} =\max\left(0, \mathsf{G}_{n-1}\right),\qquad
  \mathsf{G}_n+\mathsf{G}_{n-1} =A+2Q+2n Q-\max\left(0, n Q-\mathsf{F}_{n}\right).
\end{gather*}
\end{theorem}

\begin{proof}
We can obtain
\begin{gather*}
\mathsf{F}_n =\max\left\{F_{n+1}+\left(n+1\right)Q+A-G_n, 0\right\}
-\max\left(F_{n+1}, G_n-n  Q\right)\\
\phantom{\mathsf{F}_n}{}  =n  Q-G_n+\max\left\{A+\left(n+1\right)Q+\max\left(0, G_n\right), F_n+G_n\right\}\\
\phantom{\mathsf{F}_n=}{} -\max\left\{n  Q+\max\left(0, G_n\right), F_n+G_n\right\},
\\
\mathsf{G}_n =Q+\max\left\{\left(n+1\right)Q+A, G_n-F_{n+1}\right\}
-\max\left(F_{n+1}, G_n-n  Q\right)\\
\phantom{\mathsf{G}_n}{}
=\left(n+1\right)Q+F_n-\max\left(0, G_n\right)+\max\left\{A+\left(n+1\right)Q+\max\left(0, G_n\right),  F_n+G_n\right\}\\
\phantom{\mathsf{G}_n =}{}
-\max\left\{n  Q+\max\left(0, G_n\right),  F_n+G_n\right\}
\end{gather*}
by using \eqref{eq:udpa6f}, and
\begin{gather*}
\mathsf{F}_{n-1} =\max\left(F_{n}+n  Q+A-G_{n-1},  0\right)
-\max\left\{F_{n},  G_{n-1}-\left(n-1\right)Q\right\}\\
\phantom{\mathsf{F}_{n-1}}{} =G_n-n  Q+\max\left(F_n,  n  Q\right)-F_n
+\max\left\{G_n+\max\left(F_n,  n  Q\right),  n  Q\right\}\\
\phantom{\mathsf{F}_{n-1}=}{} -\max\left\{G_n+\max\left(F_n,  n  Q\right),  A+\left(n+1\right)Q\right\},
\\
\mathsf{G}_{n-1} =Q+\max\left\{n  Q+A,  G_{n-1}-F_{n}\right\}
-\max\left\{F_{n},  G_{n-1}-\left(n-1\right)Q\right\}\\
\phantom{\mathsf{G}_{n-1}}{} =A+\left(n+1\right)Q-F_n+\max\left\{G_n+\max\left(F_n,  n  Q\right),  n  Q\right\}\\
\phantom{\mathsf{G}_{n-1}=}{} -\max\left\{G_n+\max\left(F_n,  n  Q\right),  A+\left(n+1\right)Q\right\}
\end{gather*}
by using \eqref{eq:udpa6g}. Thus we f\/ind
\begin{gather*}
\mathsf{F}_n+\mathsf{F}_{n-1} =\max\left(0,  \mathsf{G}_{n-1}\right) \\
\phantom{\mathsf{F}_n+\mathsf{F}_{n-1}}{}
 =
\max\left(F_n,  n  Q\right)-F_n+\max\left\{A+\left(n+1\right)Q+\max\left(0, G_n\right),  F_n+G_n\right\}\\
\phantom{\mathsf{F}_n+\mathsf{F}_{n-1} =}{}
-\max\left\{G_n+\max\left(F_n,  n  Q\right),  A+\left(n+1\right)Q\right\},
\\
\mathsf{G}_n+\mathsf{G}_{n-1}
 =A+2Q+2n Q-\max\left(0, n Q-\mathsf{F}_{n}\right)
  =A+\left(2n+2\right)Q-\max\left(0, G_n\right)\\
\phantom{\mathsf{G}_n+\mathsf{G}_{n-1}=}{}
+\max\left\{A+\left(n+1\right)Q+\max\left(0, G_n\right),  F_n+G_n\right\}\\
\phantom{\mathsf{G}_n+\mathsf{G}_{n-1}=}{}-\max\left\{G_n+\max\left(F_n,  n  Q\right),  A+\left(n+1\right)Q\right\}.\tag*{\qed}
\end{gather*}
\renewcommand{\qed}{}
\end{proof}

\begin{theorem}\label{thm:bt2}
If $F_{n}$ and
$G_n$ satisfy ud-$P(A_6)$ \eqref{eqs:udpa6}, then
\begin{gather*}
\mathtt{F}_{n+1} =\max\left(n  Q+A+F_n-G_n,  0\right)
-\max\left\{F_n,  G_n-\left(n+1\right)Q\right\},\\
\mathtt{G}_{n} =-Q+\max\left(n  Q+A,  G_n-F_n\right)
-\max\left\{F_n,  G_n-\left(n+1\right)Q\right\}
\end{gather*}
satisfy ud-$P(A_6)$:
\begin{gather*}
  \mathtt{F}_n+\mathtt{F}_{n-1} =\max\left(0, \mathtt{G}_{n-1}\right),\qquad
  \mathtt{G}_n+\mathtt{G}_{n-1} =A-2Q+2n Q-\max\left(0, n Q-\mathtt{F}_{n}\right).
  \end{gather*}
\end{theorem}

\begin{proof}
We can obtain
\begin{gather*}
\mathtt{F}_{n-1} =\max\left\{\left(n-2\right)Q+A+F_{n-2}-G_{n-2},  0\right\}
-\max\left\{F_{n-2},  G_{n-2}-\left(n-1\right)Q\right\},\\
\phantom{\mathtt{F}_{n-1}}{} =\left(n-1\right)Q-G_{n-2}+\max\left\{A+\left(n-2\right)Q+\max\left(0,  G_{n-2}\right),  F_{n-1}+G_{n-2}\right\}\\
\phantom{\mathtt{F}_{n-1}=}{} -\max\left\{\left(n-1\right)Q+\max\left(0,  G_{n-2}\right),  F_{n-1}+G_{n-2}\right\}
\end{gather*}
by using \eqref{eq:udpa6f}, and
\begin{gather*}
\mathtt{F}_{n} =\max\left\{\left(n-1\right)Q+A+F_{n-1}-G_{n-1},  0\right\}
-\max\left(F_{n-1},  G_{n-1}-n  Q\right) \\
\phantom{\mathtt{F}_{n}}{} =G_{n-2}-\left(n-1\right)Q+\max\left\{F_{n-1},  \left(n-1\right)Q\right\}-F_{n-1}\\
\phantom{\mathtt{F}_{n}=}{}+\max\left[G_{n-2}+\max\left\{F_{n-1},  \left(n-1\right)Q\right\},  \left(n-1\right)Q\right]\\
\phantom{\mathtt{F}_{n}=}{}-\max\left[G_{n-2}+\max\left\{F_{n-1},  \left(n-1\right)Q\right\},  A+\left(n-2\right)Q\right],
\\
\mathtt{G}_{n-1} =-Q+\max\left\{\left(n-1\right)Q+A,  G_{n-1}-F_{n-1}\right\}
-\max\left(F_{n-1},  G_{n-1}-n  Q\right)\\
\phantom{\mathtt{G}_{n-1}}{} =A+\left(n-2\right)Q-F_{n-1}+\max\left[G_{n-2}+\max\left\{F_{n-1},  \left(n-1\right)Q\right\},  \left(n-1\right)Q\right]\\
\phantom{\mathtt{G}_{n-1}=}{}-\max\left[G_{n-2}+\max\left\{F_{n-1},  \left(n-1\right)Q\right\},  A+\left(n-2\right)Q\right]
\end{gather*}
by using \eqref{eq:udpa6g}. Thus we f\/ind
\begin{gather*}
  \mathtt{F}_n+\mathtt{F}_{n-1} =\max\left(0, \mathtt{G}_{n-1}\right)
 =\max\left\{F_{n-1},  \left(n-1\right)Q\right\}-F_{n-1}\\
\phantom{\mathtt{F}_n+\mathtt{F}_{n-1} =}{} +\max\left\{A+\left(n-2\right)Q+\max\left(0,  G_{n-2}\right),  F_{n-1}+G_{n-2}\right\}\\
\phantom{\mathtt{F}_n+\mathtt{F}_{n-1} =}{}-\max\left[G_{n-2}+\max\left\{F_{n-1},  \left(n-1\right)Q\right\},  A+\left(n-2\right)Q\right].
\end{gather*}
We obtain
\begin{gather*}
\mathtt{F}_{n} =\max\left\{\left(n-1\right)Q+A+F_{n-1}-G_{n-1},  0\right\}
-\max\left(F_{n-1},  G_{n-1}-n  Q\right) \\
\phantom{\mathtt{F}_{n}}{} =n  Q-G_{n-1}+\max\left\{A+\left(n-1\right)Q+\max\left(0, G_{n-1}\right),  F_{n}+G_{n-1}\right\}\\
\phantom{\mathtt{F}_{n}=}{}-\max\left\{n  Q+\max\left(0, G_{n-1}\right),  F_{n}+G_{n-1}\right\},
\\
\mathtt{G}_{n-1} =-Q+\max\left\{\left(n-1\right)Q+A,  G_{n-1}-F_{n-1}\right\}
-\max\left(F_{n-1},  G_{n-1}-n  Q\right)\\
\phantom{\mathtt{G}_{n-1}}{} =\left(n-1\right)Q+F_n-\max\left(0, G_{n-1}\right)\\
\phantom{\mathtt{G}_{n-1} =}{}
+\max\left\{A+\left(n-1\right)Q+\max\left(0,  G_{n-1}\right),  F_n+G_{n-1}\right\}\\
\phantom{\mathtt{G}_{n-1} =}{}
-\max\left\{n  Q+\max\left(0,  G_{n-1}\right),  F_n+G_{n-1}\right\}
\end{gather*}
by using \eqref{eq:udpa6f}, and
\begin{gather*}
\mathtt{G}_{n} =-Q+\max\left(n  Q+A,  G_{n}-F_{n}\right)
-\max\left\{F_{n},  G_{n}-\left(n+1\right)Q\right\}\\
\phantom{\mathtt{G}_{n}}{} =A+\left(n-1\right)Q-F_{n}+\max\left\{G_{n-1}+\max\left(F_{n}, n  Q\right),  n  Q\right\}\\
\phantom{\mathtt{G}_{n}=}{} -\max\left\{G_{n-1}+\max\left(F_{n},  n  Q\right),  A+\left(n-1\right)Q\right\}
\end{gather*}
by using \eqref{eq:udpa6g}. Thus we f\/ind
\begin{gather*}
\mathtt{G}_n+\mathtt{G}_{n-1}
 =A-2Q+2n Q-\max\left(0, n Q-\mathtt{F}_{n}\right)  =A+\left(2n-2\right)Q-\max\left(0, G_{n-1}\right)\\
\phantom{\mathtt{G}_n+\mathtt{G}_{n-1}=}{}
+\max\left\{A+\left(n-1\right)Q+\max\left(0, G_{n-1}\right),  F_n+G_{n-1}\right\}
\\
\phantom{\mathtt{G}_n+\mathtt{G}_{n-1}=}{}
-\max\left\{G_{n-1}+\max\left(F_n,  n  Q\right),  A+\left(n-1\right)Q\right\}.\tag*{\qed}
\end{gather*}
\renewcommand{\qed}{}
\end{proof}
So the exact solutions also can be obtained from the solution in Theorem~\ref{thm:0} by using the B\"acklund transformation.

\section{Special solutions}\label{sec:spe}

In \cite{HKW}, Hamamoto, Kajiwara and Witte constructed hypergeometric
solutions to $q$-$P(A_6)$ by applying B\"acklund transformations to the ``seed''
solution which satisf\/ies a Riccati equation.
Their solutions have a determinantal
form with basic hypergeometric function elements whose continuous
limits are showed by them to be Airy functions, the hypergeometric solutions
of the second Painlev\'e equation.
In \cite{NishiokaS,NishiokaS2}, S.~Nishioka proved that transcendental solutions of~\mbox{$q$-$P(A_6)$} in a decomposable
extension may exist only for special parameters, and that each of
them satisf\/ies the Riccati equation mentioned above if we apply the B\"acklund
transformations to it appropriate times.
He also proved non-existence of algebraic solutions.

$q$-$P(A_6)$ \eqref{eqs:qpa6} for $a=q^{2m+1}\ (m\in \mathbb{Z})$ has the hypergeometric solution.
The case of $A=\left(2m+1\right)Q$ in ud-$P(A_6)$ corresponds to $a=q^{2m+1}$ in the discrete system.
It is hard to apply the ultradiscretization procedure to the hypergeometric series.
However according to~\cite{Ormerod}, an ultradiscrete hypergeometric solution is given in terms of $n  Q$ and $\left(-1\right)^n Q$.
If $h_1=h_2=0$ and $r=1/2$ in Theorem~\ref{thm:0}, then we obtain an ultradiscrete hypergeometric solution of ud-$P(A_6)$ for $A=Q$:
\begin{gather*}
F_n =
\begin{cases}
\frac{1}{3}Q\left(-1\right)^{n}&(n\le -1),\\
\frac{n+1}{3}Q&(n\ge 0),
\end{cases}\qquad
G_n =
\begin{cases}
\left(n+1\right)Q&( n\le -1),\\
\frac{2n+3}{3}Q&(n\ge 0).
\end{cases}
  \end{gather*}
If $h_1=h_2=0$ and $r=1/2$ in Theorem~\ref{thm:1}, then we obtain an ultradiscrete hypergeometric solution of ud-$P(A_6)$ for $A=\left(2m+1\right)Q$ $(m\in \mathbb{N})$:
\begin{gather*}
F_n =
\begin{cases}
\frac{1}{3}Q\left(-1\right)^{n+m}&(n\le -m-1),\\
\frac{2n+2m+1}{4}Q+\frac{1}{12}Q\left(-1\right)^{n-m}&(-m\le n\le m-1),\\
\frac{n+2m+1}{3}Q&(n\ge m),
\end{cases}\\
G_n =
\begin{cases}
\left(n+m+1\right)Q&( n\le m-1),\\
\frac{2n+4m+3}{3}Q&(n\ge m).
\end{cases}
  \end{gather*}
If $h_1=h_2=0$ and $r=1/2$ in Theorem~\ref{thm:2}, then we have an ultradiscrete hypergeometric solution for $A=(2m+1)Q$ $(-m\in \mathbb{N})$:
\begin{gather*}
F_n =
\begin{cases}
0&(n\le -2m-1),\\
\frac{n+2m+1}{3}Q&(n\ge -2m),
\end{cases}\\
G_n =
\begin{cases}
\left(n+m+1\right)Q&( n\le -1),\\
\frac{2n+4m+3}{4}Q-\frac{1}{12}Q\left(-1\right)^n&(0\le  n\le -2m-1),\\
\frac{2n+4m+3}{3}Q&(n\ge -2m).
\end{cases}
  \end{gather*}

\section{Concluding remarks}\label{sec:rem}
We have given the ultradiscrete analogue of $q$-$P(A_6)$.
Moreover, we have presented the exact solutions with two parameters.
These solutions are expressed by using linear functions and periodic functions.
But the exact solution is only useful when the two parameters are in
a~limited range.
If one wants to construct the exact solution for any initial values,
then one needs to use a~multitude of branches with respect to $n$ in order to express a solution.
We have also presented its special solutions that correspond to the hypergeometric solutions of~\mbox{$q$-$P(A_6)$}.
The ultra\-discrete hypergeometric solutions are included in the resulting solutions with two parameters.

There are many studies on analytic properties of solutions to the Painlev\'e equations~\cite{Doyon,DM,Jimbo}.
But there exist few studies on analytic properties of the $q$-Painlev\'e equations~\cite{Mano,Ohyama}.
We hope to study the $q$-Painlev\'e equations by employing the results in the ultradiscrete systems.

\pdfbookmark[1]{References}{ref}
\LastPageEnding

\end{document}